\documentclass[USletter, 10 pt, conference]{IEEEtran}

\usepackage{cite}
\usepackage{amsmath,amssymb,amsfonts}

\usepackage{graphicx}
\usepackage{textcomp}
\usepackage{xcolor}
\usepackage{verbatim}
\usepackage{booktabs}
\usepackage{graphicx,array,xcolor,colortbl}
\usepackage{pifont}
\usepackage{pifont}     
\newcommand{\cmark}{\ding{51}} 
\newcommand{\xmark}{\ding{55}} 

\newcommand{\red}[1]{#1}

\definecolor{headgray}{gray}{0.90}
\def\BibTeX{{\rm B\kern-.05em{\sc i\kern-.025em b}\kern-.08em
    T\kern-.1667em\lower.7ex\hbox{E}\kern-.125emX}}
\usepackage{float}

\usepackage{titlesec}

\usepackage{tabularx}   
\usepackage{booktabs}   
\usepackage{array}      
\usepackage{comment}
\usepackage{xcolor}
\usepackage{amsmath,amssymb}
\usepackage{mathtools} 
\usepackage{subcaption}
\usepackage{graphicx}

\usepackage{booktabs}
\usepackage{array}
\usepackage{tabularx}   
\newcolumntype{Y}{>{\raggedright\arraybackslash}X} 

\usepackage{siunitx}
\usepackage{enumitem} 

\usepackage{algorithm}
\usepackage{algpseudocode} 

\usepackage{tikz}
\usetikzlibrary{arrows.meta,positioning,shapes.geometric,fit}

\usepackage{xurl}                 
\usepackage[hidelinks]{hyperref}  

\usepackage[nameinlink,capitalize,noabbrev]{cleveref} 

\usepackage{cite} 

\usepackage{amsmath,amssymb}
\usepackage{array,tabularx,booktabs}
\usepackage{float} 
\usepackage{array,makecell,tabularx,booktabs}

\newcolumntype{L}{>{\raggedright\arraybackslash}p{.40\columnwidth}}
\newcolumntype{C}{>{\centering\arraybackslash}p{.10\columnwidth}}

\usepackage{xcolor}

\IEEEoverridecommandlockouts                              

\begin{document}
\title{DexterousMag: A Reconfigurable Electromagnetic Actuation System for Miniature Helical Robot
}

\author{Jialin Lin, Dandan Zhang
\thanks{Jialin Lin, Dandan Zhang are with the Department of Bioengineering, Imperial College London.}
}

\maketitle

\begin{abstract}
\red{
Despite the promise of magnetically actuated miniature helical robots for minimally invasive interventions, state-of-the-art electromagnetic actuation systems are often space-inefficient and geometrically fixed. These constraints hinder clinical translation and, moreover, prevent task-adaptive trade-offs among workspace coverage, energy distribution, and field/gradient capability. We present DexterousMag, a robot-arm-assisted three-coil electromagnetic actuation system that enables continuous geometric reconfiguration of a compact coil group, thereby redistributing magnetic-field and gradient capability for task-adaptive operation. The reconfiguration is realized by a parallel mechanism that exposes a single geometric DOF of the coil group, conveniently parameterized by the polar angle $\theta$. Using an FEM-based modeling pipeline, we precompute actuation and gradient libraries and quantify the resulting trade-offs under current limits: configurations that favor depth reach expand the feasible region but reduce peak field/gradient, whereas configurations that favor near-surface capability concentrate stronger fields/gradients and support lifting. We validate these trade-offs on representative tasks (deep translation, planar tracking, and 3D lifting) and further demonstrate a proof-of-concept online geometry scheduling scheme for combined tasks, benchmarked against fixed-geometry settings. Overall, DexterousMag establishes continuous geometric reconfiguration as an operational mechanism for enlarging the practical envelope of miniature helical robot actuation while improving energy efficiency and safety. }
\end{abstract}


\section{Introduction}
\red{
Magnetic actuation is a compelling approach for driving small-scale robots in biomedical settings because it is wireless, untethered, and can provide precise remote control within enclosed and delicate environments~\cite{lin_magnetic_2024}. By enabling targeted locomotion and positioning, magnetically actuated robots can, in principle, deliver therapeutic agents directly to diseased tissues, reducing systemic dispersion and off-target effects, and improving access to hard-to-reach sites~\cite{alapan_multifunctional_2020}. These advantages motivate continued development of magnetic actuation platforms that can support long-range navigation and task-adaptive control under realistic anatomical and procedural constraints.}
\begin{figure}
    \centering
     \captionsetup{font=footnotesize,labelsep=period}
    \includegraphics[width=1\linewidth]{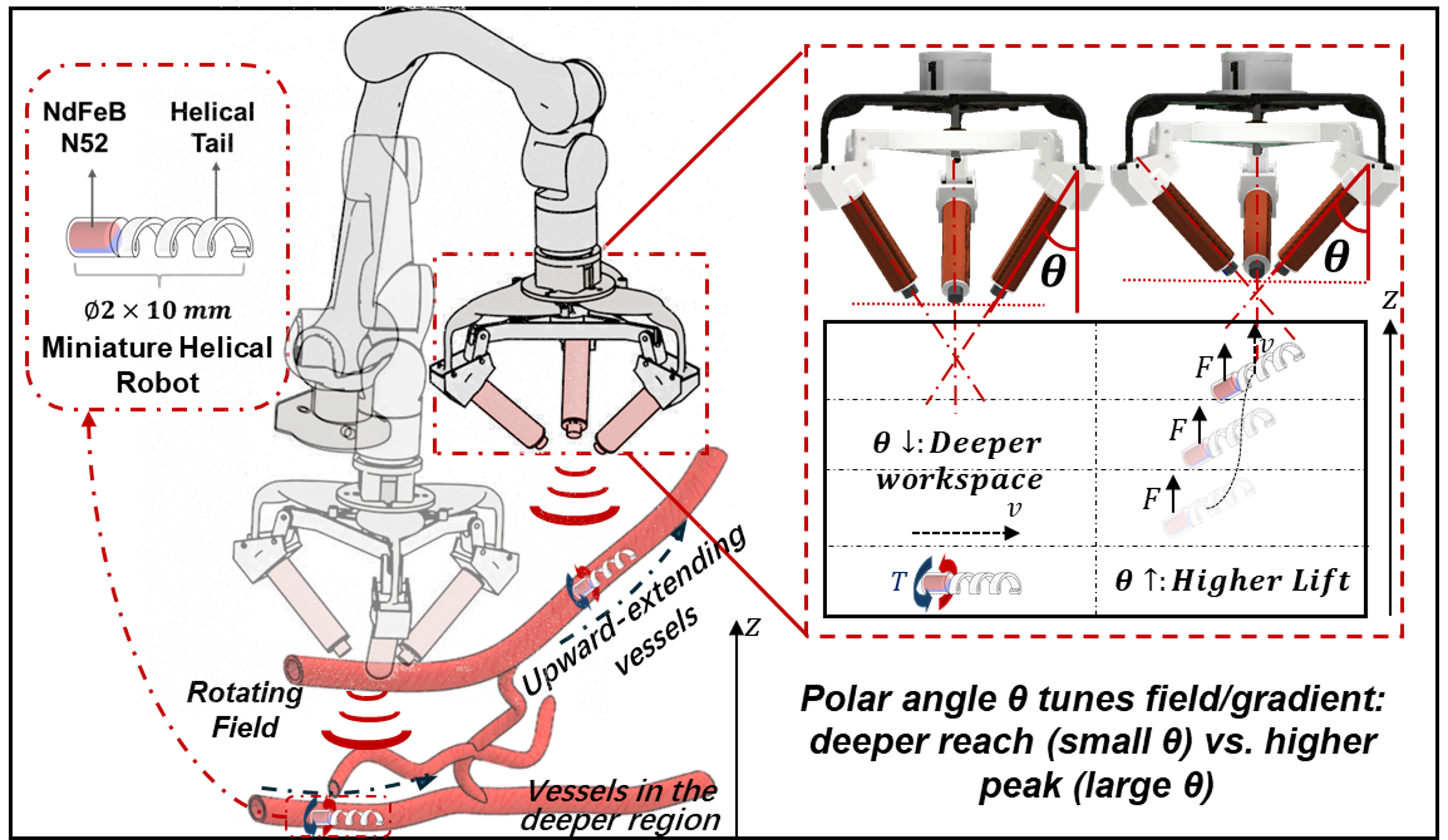}
\caption{Concept of DexterousMag: a reconfigurable, robot-arm-assisted electromagnetic actuation system that redistributes magnetic energy and gradient capability through continuous coil-group reconfiguration (parameterized by $\theta$)  for task-adaptive control of a miniature helical robot for biomedical treatment.}

    \label{fig:concept}
\end{figure}
\red{
Existing magnetic actuation platforms can be organized along two orthogonal axes: field source (permanent-magnet vs.\ electromagnetic) and deployment strategy (stationary vs.\ robot-arm-assisted). Permanent-magnet systems exploit strong static fields and spatial gradients to pull or orient miniature robots, often by translating or rotating magnets~\cite{nadour_cochlerob_2023,mao2025deep, fountain_wireless_2010}. While their high field strength is attractive, permanent magnets cannot be instantaneously switched off, which makes transient forces difficult to ``hard-stop'' and complicates fine force modulation near delicate vessel walls~\cite{shao_study_2021}. This limitation is particularly relevant for safety-critical interactions and closed-loop control where rapid field shutoff can be desirable.
Electromagnetic actuation systems (EMAs), by contrast, can modulate and null magnetic fields/gradients by controlling coil currents, offering stronger controllability and potentially safer operation. Traditional stationary EMAs, as summarized in Table \ref{tab:compare_sota},  typically employ fixed coil arrangements, including Helmholtz pairs~\cite{wu_3-d_2020}, Maxwell coils~\cite{song_electromagnetic_2017}, and multi-coil platforms such as OctoMag~\cite{kummer_octomag_2010}, CardioMag~\cite{charreyron_modeling_2021}, and Navion~\cite{gervasoni_humanscale_nodate}. These systems can generate uniform fields and/or controllable gradients with good isotropy in a bounded region, supporting high-precision manipulation. However, their fixed geometries often translate into poor space utilization and a confined feasible workspace, limiting practical in vivo navigation over long access paths from distal entry points to deep targets. In clinical workflows, this limitation commonly leads to hybrid strategies (e.g., catheterization to place an agent near the target before release)~\cite{nelson_delivering_2023}, which forfeit end-to-end magnetic navigation and reintroduce catheter-related risks and procedural complexity.}

\begin{table*}[t]
\centering
\caption{Comparison with representative magnetic actuation systems. Checks indicate features explicitly reported.}
\label{tab:compare_sota}
\resizebox{\textwidth}{!}{
\begin{tabular}{lcccccccc}
\toprule
\textbf{System} & \textbf{Mounting} & \textbf{\# Coils} & \textbf{Repositioning} & \textbf{Tunable $\theta$} & \textbf{$\theta\!\to\!$Workspace/Gradient/Energy Model} & \textbf{Space efficiency} & \textbf{Validated Task(s)} \\
\midrule
OctoMag~\cite{kummer_octomag_2010} & Stationary & 8     & \xmark & \xmark & \xmark & \textit{Poor} & Local interventions \\
CardioMag/Navion~\cite{song_electromagnetic_2017,charreyron_modeling_2021} & Stationary & Multi  & \xmark & \xmark & \xmark & \textit{Poor} & Local interventions \\
Human-scale EMA~\cite{go_human_2020,gervasoni_humanscale_nodate} & Stationary & Multi  & \xmark & \xmark & \xmark & \textit{Poor} & Local interventions \\
DeltaMag~\cite{yang_deltamag_2019} & Parallel robot & 3    & \cmark & \xmark      & \xmark & \textit{Moderate (planar)} & 2D tube motion \\
RoboMag~\cite{du_robomag_2020,cai_performance-guided_2024} & 3 Arms & 3        & \cmark & \xmark      & \xmark & \textit{Better (limited by 3-arm intersection)} & 3D helical robot \\
TrinityMag~\cite{li_trinitymag_2024} & Single Arm  & 3     & \cmark & \xmark     & \xmark & \textbf{Maximised} & 2D milli-robot \\
\midrule
\textbf{DexterousMag (this work)} & \textbf{Single Arm} & \textbf{3} & \textbf{\cmark} & \textbf{\cmark} & \textbf{\cmark} & \textbf{Maximised} & \textbf{Deep reach, cruise \& near-surface lift} \\
\bottomrule
\end{tabular}}
\vspace{2mm}
\end{table*}

\red{To alleviate geometric constraints and enlarge coverage, arm-assisted designs move coils relative to the target.  DeltaMag~\cite{yang_deltamag_2019} employs a parallel mechanism and demonstrates 2D tubular locomotion of helical robots. RoboMag~\cite{du_robomag_2020,cai_performance-guided_2024} uses three robotic arms carrying coils to generate magnetic torque and gradient forces in 3D, reporting a substantially larger hemispherical workspace, though the effective region remains constrained by arm intersection and mechanical reach. TrinityMag~\cite{li_trinitymag_2024} mounts three coils on a single robotic arm to navigate a cubic millirobot in 2D. Collectively, these works primarily expand workspace by repositioning coils, but they do not provide a unified framework that quantifies how the intrinsic coil-group geometry governs the redistribution of field/gradient capability under current limits, nor do they experimentally validate such geometric tuning as a task-adaptive operational parameter for 3D miniature helical robot actuation in viscous media.}

\red{Motivated by this gap, we argue (\ Fig.~\ref{fig:concept})  the coil-group geometry should be treated as an operational variable. In DexterousMag, this geometry is  parameterized by $\theta$. Changing~$\theta$ directly redistributes magnetic energy in space and therefore reshapes the achievable field magnitude and gradient distribution under a fixed current budget. Intuitively, decreasing~$\theta$ spreads the feasible actuation region deeper into the workspace but lowers peak field/gradient. Increasing~$\theta$ concentrates magnetic energy closer to the coil assembly, shrinking the feasible region while boosting near-coil field/gradient. This trade-off is especially relevant for miniature helical robots, where dipole-gradient interaction can generate a vertical lifting component $F_z$ that helps counter gravity and maintain elevation in viscous media. Treating~$\theta$ as an operational knob thus enables task-adaptive trade-offs among depth reach, control accuracy, and power/current usage.}
\begin{figure*}[h]
    \centering
     \captionsetup{font=footnotesize,labelsep=period}
    \includegraphics[width=\linewidth]{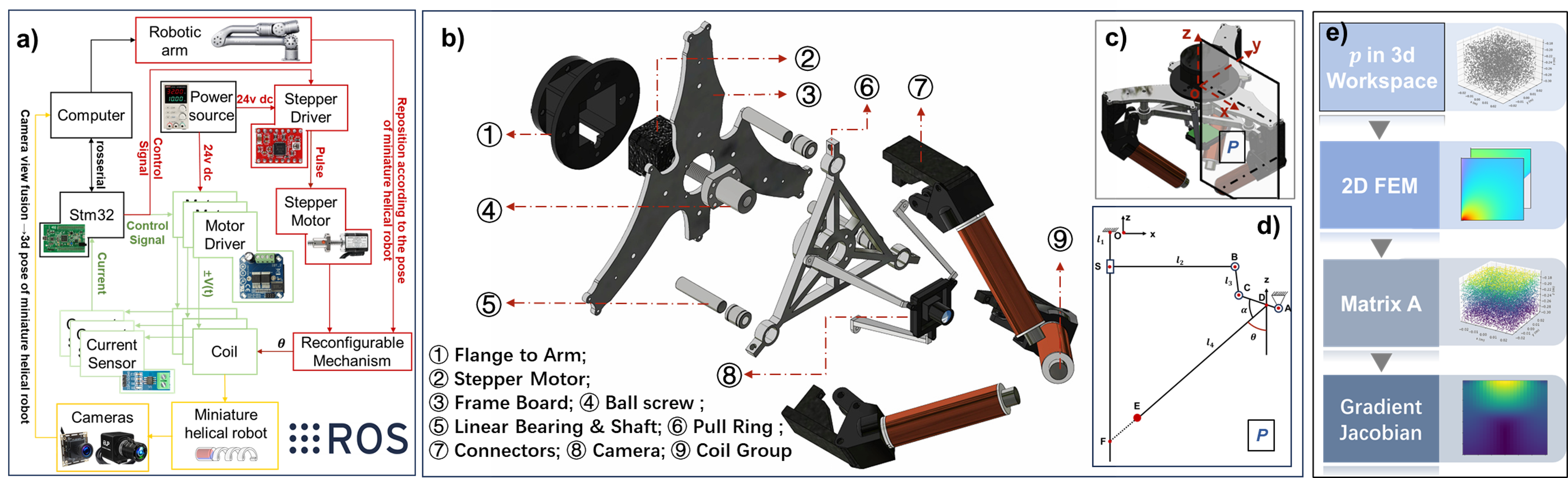}
    \caption{DexterousMag overview. 
    (a) Schematic of the hardware and signal architecture. 
    (b) Exploded view of the proposed mechanism that converts stepper–motor rotation into changes of the coil-group polar angle~$\theta$. 
    (c) Assembly diagram; plane~$P$ denotes the cross-section of one coil mechanism. 
    (d) One-coil mechanism: a planar four-bar linkage in plane~$P$. 
    (e) Modeling pipeline for building offline libraries of the actuation matrix~$\mathbf A(\mathbf p,\theta)$ and gradient Jacobian~$\mathbf {\mathcal{G} (\mathbf p,\theta)}$, enabling offline workspace analysis and online magnetic-field synthesis.}

    \label{fig:hardware}
\end{figure*}

In this paper, we present DexterousMag, a robot-arm-assisted, reconfigurable EMA that implements this geometric-tuning principle with three coils mounted on a compact parallel mechanism. Our contributions are threefold:
\begin{itemize}
    \item We present DexterousMag, a robot-arm-assisted, reconfigurable EMA that employs three coils on a compact parallel mechanism to drive small-scale helical robots and enable 3D motion in viscous media.
    \item We develop an FEM-based pipeline to quantify fields and gradients for workspace evaluation and field synthesis under current limits.
    \item \red{We experimentally validate task-adaptive reconfiguration: varying the coil-group geometry (\(\theta\)) trades depth reach for near-surface field/gradient concentration and lift \(F_z\), and we further demonstrate a proof-of-concept online \(\theta(z)\) scheduling scheme on a combined-task trajectory, benchmarked against fixed-\(\theta\) settings.} 

\end{itemize}


 



\section{Method}
In this section, we firstly define frames and conventions globally. We then present the hardware–software overview of \textit{DexterousMag} (Fig.~\ref{fig:hardware}a), detail the reconfigurable coil mechanism (Fig.~\ref{fig:hardware}b-d), and finally describe the modeling pipeline used to build offline libraries of the actuation matrix and gradient Jacobian. (Fig.~\ref{fig:hardware}e). This pipeline is used to test how the polar angle~$\theta$ reshapes the field/gradient distribution under current limits, in order to select task-oriented $\theta$.

\subsection{Frames and Conventions}
The coil assembly is threefold rotationally symmetric about the C\(_3\) axis. We use two right-handed frames: the world frame \(\mathcal F_w\), which is fixed to the table and coincident with the robotic arm’s base frame \texttt{link\_00}, and the mechanism frame \(\mathcal F_m\)  (Fig.~\ref{fig:hardware}c), which is rigidly attached to the coil assembly with its origin on the C\(_3\) axis. Because the mechanism is mounted on the arm, it can be treated as the arm’s end effector, and unless stated otherwise all fields, coordinates, and plots are expressed in \(\mathcal F_m\). For brevity, we write \(\mathbf p \equiv \mathbf p_m = [x\ y\ z]^{\top}\) and \(z \equiv z_m\); “deeper” denotes motion toward more negative \(z\) along the C\(_3\) axis of \(\mathcal F_m\).

\subsection{System Overview}

\paragraph*{Hardware}
DexterousMag consists of a host PC, an STM32-based current-control unit, a 6-DoF robotic arm, and a three-coil electromagnetic actuation module mounted at the arm flange. Each coil is driven by an H-bridge with inline current sensing. A stepper-driven ball-screw stage actuates a compact mechanism that synchronously adjusts the coil-group polar angle $\theta$. \red{Perception is provided by an eye-in-hand camera aligned with the coil triad centreline and a fixed side-view camera; together they enable stereo 3-D pose estimation of the miniature helical robot. This configuration mirrors the intended in vivo workflow, where the arm carries local imaging (e.g., ultrasound).} Coil specifications are reported on the project website (Table~S1), and the hardware connections are summarized in Fig.~\ref{fig:hardware}a.

\paragraph*{Software}
All software runs in a Dockerised Ubuntu/ROS~Noetic environment for  reproducibility. The microcontroller interfaces with ROS via rosserial. The cameras are calibrated and registered to the robot using standard hand-eye calibration, and the robot description publishes the TF tree for frame transformations. The robot 3-D pose is estimated from principal-axis recovery and plane intersection~\cite{xu_planar_2015}. During experiments, the arm performs visual servoing to maintain the robot near the image center while aligning it with the C3 axis.  

\subsection{Reconfigurable Coil Mechanism}

Robot-arm-assisted EMA arrays can tune the azimuthal interval $\phi$, the polar angle $\theta$, and the target-coil spacing $r$~\cite{cai_performance-guided_2024}. For a tri-coil layout, $\phi=120^\circ$ provides near-isotropic fields and $r$ is typically adjusted by arm positioning; therefore, we treat $\theta$ as the primary reconfiguration variable that redistributes field energy, trading depth reach against near-surface field strength and gradients.

\paragraph*{Architecture.}
Fig.~\ref{fig:hardware}b-c illustrates the proposed mechanism. Three identical coplanar PRRR four-bar units are coupled by a common pull ring, such that a single prismatic input produces a synchronized change in the polar angle $\theta$ across all three coils.\red{The current mechanism couples the three coils to preserve the coil group's \(C_3\) symmetry, exposing a single interpretable reconfiguration DOF \(\theta\).} A stepper motor drives a ball screw to translate the pull ring along the vertical axis, while linear guides constrain the motion and bearings provide low-friction revolute joints. A sacrificial 3D-printed flange is used for collision protection, and the lightweight assembly remains within the arm payload limit.

\paragraph*{Planar kinematics.}
For kinematic derivation, we define a cross-sectional plane \(P\) that contains one coil unit (Fig.~\ref{fig:hardware}c); the same plane is sketched in Fig.~\ref{fig:hardware}d to introduce the planar geometry. 
The point $O$ coincides with the origin of frame \(\mathcal F_m\).
This 2-D slice captures the PRRR loop and the \(s\!\to\!\theta\) mapping used for mechanism actuation.
Let the slider $S$ translate along $OS$ with travel \(s=\lvert OS\rvert\). Denote the revolute chain $B\!\to\!C\!\to\!A$ and let $D$ be a point on link $CA$ (Fig.~\ref{fig:hardware} d); $A$ is the ground. The feasible coupler position $C$ satisfies
\[
\|C-B\|=\lvert BC\rvert,\qquad \|C-D\|=\lvert CD\rvert .
\]
Define the rocker orientation from the $+x$ axis by
\[
\psi=\operatorname{atan2}(z_C-z_D,\,x_C-x_D).
\]
The coil axis $DE$ is obtained by rotating $\overrightarrow{DC}$ counter-clockwise by a fixed offset $\alpha=\pi/3$, i.e. $\varphi=\psi+\alpha .$
With $+z$ pointing upward, the coil polar angle and the coil-tip position are
\[
\;\theta=\bigl|\varphi-\tfrac{\pi}{2}\bigr|,\qquad
E=D+\lvert DE\rvert\,[\,\cos\varphi,\,\sin\varphi\,].\;
\]
This kinematic mapping $s \mapsto \theta$ is used to command the stepper stage and to register $\theta$ for simulation analysis and real platform experiment. 

\begin{figure*}
    \centering
     \captionsetup{font=footnotesize,labelsep=period}
    \includegraphics[width=0.95\linewidth]{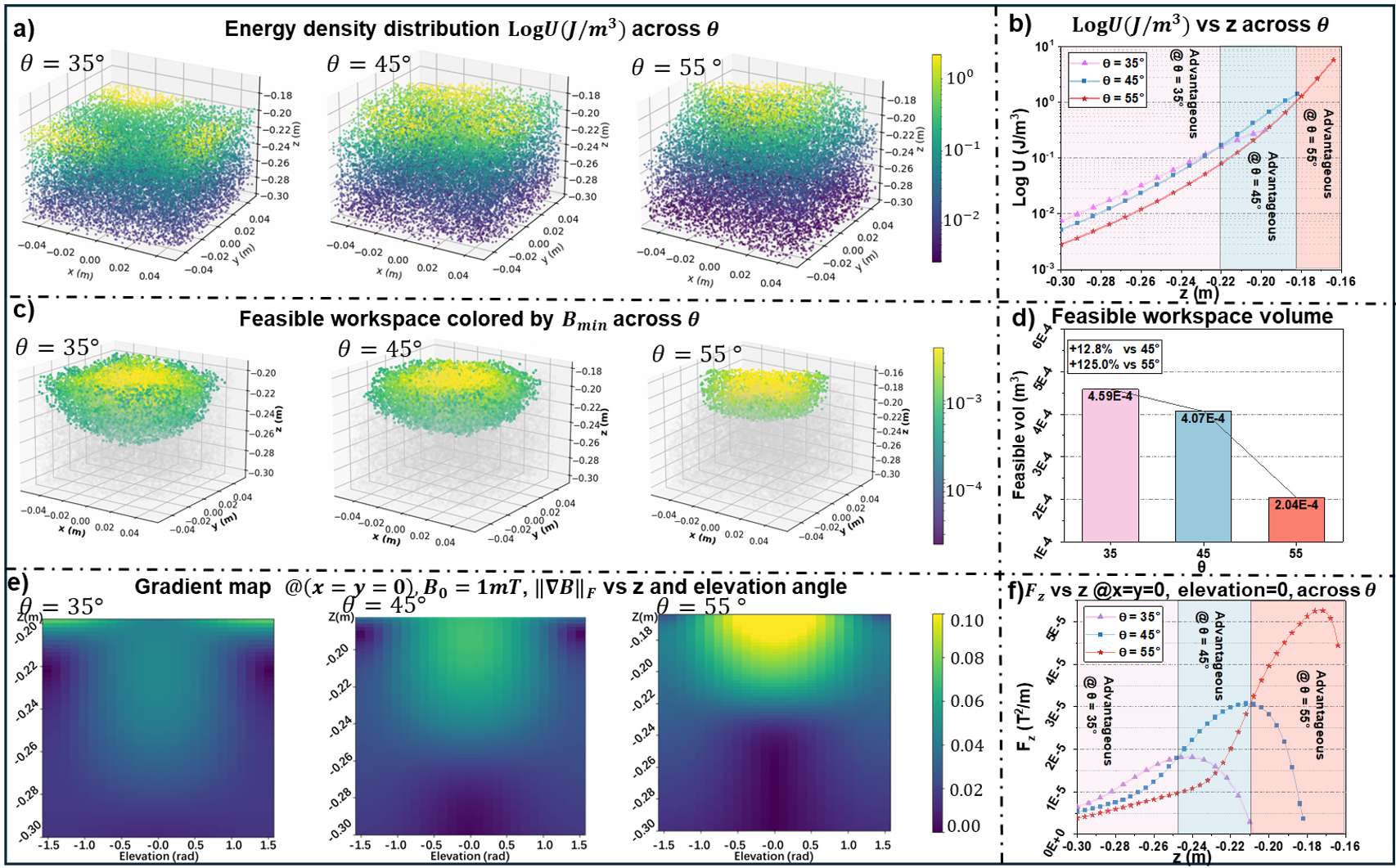}
    \caption{Simulated workspace analysis across \(\theta\).
  (a) Energy density \(\log u\) at \(\theta=35^\circ,45^\circ,55^\circ\).
  (b) Mean \(\log u\) vs.\ depth \(z\) (at \(x{=}y{=}0\)).
  (c) Feasible workspace \(\mathcal W_\theta\) colored by \(B_{\min}\).
  (d) Effective workspace volume \(V(\theta)\) (convex hull).
  (e) Gradient disturbance map \(g_F{=}\|\nabla\mathbf B\|_F\) vs.\ \(z\) and elevation \(\varepsilon\) under \(B_0{=}1\) mT; azimuth is averaged due to threefold symmetry.
  (f) Cycle-averaged vertical force \(F_z\) vs.\ \(z\) at \((x,y)=(0,0)\), \(\varepsilon{=}0\).
  All panels use common color scales and current limits for across \(\theta\) comparability.
  }
    \label{fig:simulation res}
\end{figure*}

\subsection{Actuation Matrix and Gradient Jacobian}
Unlike conventional robots whose power is transmitted through rigid links, a magnetic miniature helical robot is driven by a contactless field. Control thus begins by identifying the linear maps from coil currents to the local magnetic field and its gradients. \red{We therefore build reusable FEM-derived libraries to quantify feasibility and field/gradient authority under current limits and collision constraints. Directly calibrating \(\mathbf{A}\) and \(\mathcal{G}\) on hardware is time-consuming and costly, while full-assembly FEM at every configuration is prohibitive, especially in a reconfigurable structure where changing \(\theta\) reshapes the field/gradient envelope and shifts the collision-free workspace.}

\red{To balance accuracy and efficiency, we adopt the pipeline in Fig.~\ref{fig:hardware}e: (i) define an anti-collision workspace $\mathcal W(\theta)$ that respects geometric clearances of the arm-mechanism-coil stack; (ii) precompute single-coil FEM fields for unit current, and place each coil in pose by rigid spatial projection for any given $\theta$; and (iii) aggregate these fields to build offline libraries $\mathbf A(\mathbf p,\theta)$ and $\mathcal G(\mathbf p,\theta)$ over $\mathbf p\!\in\!\mathcal W(\theta)$ and a discrete set of $\theta$.}

\subsubsection{Definition of Anti-Collision Workspace}
\label{subsec:anti_collision_workspace}

When the coil inclination angle $\theta$ changes, the overall height of the mechanism increases. To prevent collision with the human body in envisioned in vivo scenarios and with the experimental fixture in our setup, we define
\[
\mathcal{W}(\theta)=
[-0.05,\,0.05]^2 \times [-0.3,\,z_{\mathrm{limit}}(\theta)] .
\]
Here, $x$ and $y$ span a $0.05$\,m $\times$ $0.05$\,m region to focus on field distributions as depth $z$ increases. The upper bound $z_{\text{limit}}(\theta)$ is the maximum admissible height without collision for a given $\theta$, defined in \(\mathcal F_m\).
\begin{figure*}
    \centering
     \captionsetup{font=footnotesize,labelsep=period}
    \includegraphics[width=.95\linewidth]{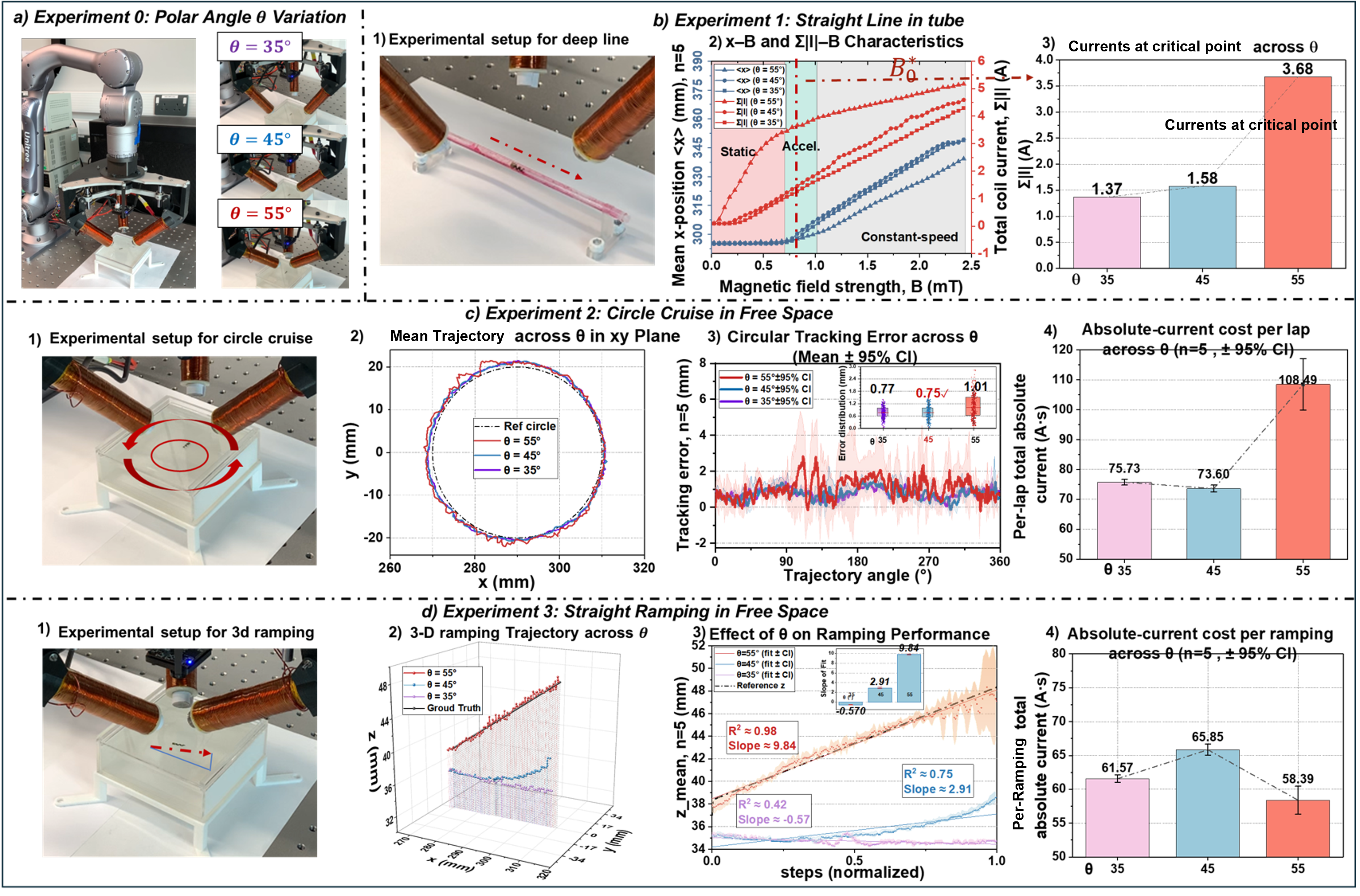}
   \caption{Experiments 1-3 on the physical DexterousMag.
a) Three mechanism configurations at \(\theta=35^\circ,45^\circ,55^\circ\).
b) Exp.~1:  Straight line in a tube. Let \(B_0^{\star}\) denote the critical field magnitude at the target point at which the miniature helical robot transitions from static to accelerated motion. We compare the total coil current required to reach \(B_0^{\star}\) across \(\theta\): smaller \(\theta\) attains the same depth with less current, indicating greater depth reach per ampere.
c) Exp.~2: Circular tracking in free space (\(B=1\,\mathrm{mT}\)): \(\theta\approx45^\circ\) yields the lowest planar tracking error with comparable current to \(35^\circ\), while \(55^\circ\) degrades tracking due to stronger, more disruptive gradients.
d) Exp.~3: Straight \(z\)-ramping in free space: \(\theta=55^\circ\) provides steeper gradients and the most efficient  lifting.}

    \label{fig:real_platform_exp}
\end{figure*}
\subsubsection{Modeling of Actuation Matrix and Gradient Jacobian}

The actuation matrix $\mathbf A(\mathbf p,\theta)$ is constructed from FEM of the individual coils. Each coil is modeled with a two-dimensional axisymmetric FEM setup at 1\,A, yielding $B_r(r,z)$ and $B_z(r,z)$; by linearity assumption, fields scale proportionally with current. For efficient evaluation, the FEM results are interpolated.

Evaluation is restricted to $\mathcal W(\theta)$, which accounts for coil geometry and mechanical clearance. For each workspace point $\mathbf p$ and inclination $\theta$, the local cylindrical coordinates $(r_i,z_i)$ relative to coil $i$ are obtained by projection $P_i(\theta)$. Interpolated FEM data provide:
\[
B_{r_i}=B_{r_i}(r_i,z_i),\qquad B_{z_i}=B_{z_i}(r_i,z_i),
\]
which are transformed to the global frame using the local radial unit vector $\mathbf e_{r_i}(\theta)$ and coil-axis direction $\mathbf u_i(\theta)$:
\[
\mathbf a_i(\mathbf p,\theta)=B_{r_i}\,\mathbf e_{r_i}(\theta)+B_{z_i}\,\mathbf u_i(\theta).
\]
Collecting the three coil contributions gives:
\[
\mathbf A(\mathbf p,\theta)=
\big[\,\mathbf a_1(\mathbf p,\theta)\ \mathbf a_2(\mathbf p,\theta)\ \mathbf a_3(\mathbf p,\theta)\,\big]
\]
where the $k$-th column $\mathbf a_k(\mathbf p,\theta)\in\mathbb R^3$ is the global field from coil $k$ at 1\,A. Hence
$\mathbf B(\mathbf p,\theta)=\mathbf A(\mathbf p,\theta)\,\mathbf i$ for any current vector $\mathbf i=[i_1,i_2,i_3]^T$.
For each coil $k$ (1\,A), define the per-coil gradient tensor
\[
\mathcal G_k(\mathbf p,\theta)\triangleq \nabla \mathbf a_k(\mathbf p,\theta)\in\mathbb R^{3\times3}.
\]
We compute $\mathcal G_k$ numerically from the interpolated single-coil FEM field using second-order central differences in $x,y,z$, with a step size $\delta$ matched to the spatial sampling resolution. \red{With three coils, we do not claim full 3D gradient controllability; instead, \(\mathcal G_k(\mathbf p,\theta)\) is used to characterize \(\theta\)-dependent parasitic gradient components within the collision-free workspace.}
Collect the three tensors as:
\[
\mathcal G(\mathbf p,\theta)\;=\;[\;\mathcal G_1(\mathbf p,\theta),\ \mathcal G_2(\mathbf p,\theta),\ \mathcal G_3(\mathbf p,\theta)\;],
\]
which we refer to as the gradient tensor basis.Then, for any current vector $\mathbf i=[i_1,i_2,i_3]^T$,
\[
\;\nabla \mathbf B(\mathbf p,\theta;\mathbf i)
\;=\;\sum_{k=1}^{3} i_k\,\mathcal G_k(\mathbf p,\theta)\;\in\mathbb R^{3\times3}. \;
\]
\red{Together with \(\mathbf B=\mathbf A\,\mathbf i\), these libraries support constraint-aware analysis of energy distribution, feasible workspace, and field/gradient (including lift) maps, and they enable quantitative comparison across reconfigurable geometries.}

\subsection{Workspace Analysis}
We analyze the FEM–derived actuation matrix $A(\mathbf{p},\theta)$ and Jacobian $\mathcal G(\mathbf p,\theta)\;$ to quantify how the reconfigurable polar angle $\theta$ redistributes magnetic energy density and thereby the feasible workspace, gradient and force. 

\subsubsection{Energy Distribution}
Magnetic energy density is
\begin{equation}
u(\mathbf{p},\theta;\mathbf{i}) \;=\; \tfrac{1}{2\mu_0}\,\|\mathbf{B}(\mathbf{p},\theta;\mathbf{i})\|_2^2,
\label{eq:energy}
\end{equation}
which scales quadratically with current. For visualization, we set $\mathbf{i}=\mathbf{1}$ (each coil at $1$\,A).

\begin{figure*}
    \centering
     \captionsetup{font=footnotesize,labelsep=period}
    \includegraphics[width=0.95\linewidth]{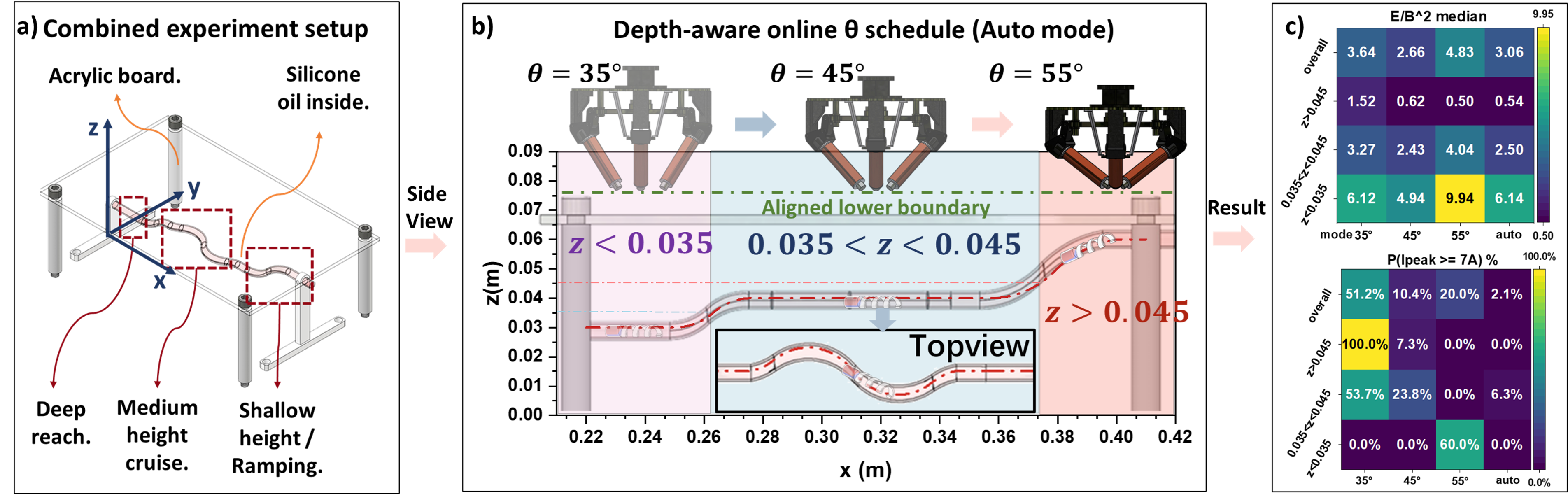}
\caption{Experiment~4: Depth-based $\theta$ scheduling in a tortuous tube under an obstacle constraint. 
(a) Setup: a 3D-printed tortuous tube provides a vessel-like path, and an acrylic plate emulates anatomical thickness and imposes a collision constraint. The trajectory concatenates the three task primitives from Experiments~1-3 (deep reach, cruise, and shallow/climb). 
(b) Tube views and the depth-triggered schedule showing online updates of $\theta$ based on target robot's height. For fair comparison, we align a common \emph{collision-free lower boundary} across configurations to decouple $\theta$-dependent geometry from the effect of field redistribution. 
(c) Results: median $E/B^2$ and $P(I_{\text{peak}}\ge 7~\mathrm{A})$ across depth-triggered auto $\theta$  schedule  and fixed 35,45,55 $\theta$  }
    \label{fig:theta_schedule_exp}
\end{figure*}
\subsubsection{Feasible Workspace Under Current Limits}
Since actuation is magnetic rather than transmitted through rigid links, the conventional manipulator notion of “workspace” does not apply. We therefore adopt the feasibility-based definition of workspace proposed by Boehler \textit{et al.}~\cite{boehler_workspace_2023}; our formulation is an adapted specialization of their method.

We evaluate feasibility under per–coil current limits $\|\mathbf{i}\|_\infty\le I_{\max}$. The maximum field along a unit direction $\mathbf{u}\in\mathbb{S}^2$ is given by the $\ell_\infty$–$\ell_1$ duality:
\begin{equation}
\max_{\|\mathbf{i}\|_\infty \le I_{\max}} \mathbf{u}^\top A(\mathbf{p},\theta)\mathbf{i}
= I_{\max}\,\|A(\mathbf{p},\theta)^\top \mathbf{u}\|_1.
\end{equation}
The worst-case directional capability is:
\begin{equation}
B_{\min}(\mathbf{p},\theta)=\min_{\|\mathbf{u}\|_2=1} I_{\max}\,\|A(\mathbf{p},\theta)^\top \mathbf{u}\|_1,
\end{equation}
approximated by quasi-uniform sphere sampling (e.g., Fibonacci grid). 
The feasible workspace for inclination $\theta$ is:
\begin{equation}
\mathcal{W}_\theta=\{\mathbf{p}:\widehat{B}_{\min}(\mathbf{p},\theta)\ge B_{\text{req}}\},
\end{equation}
and its effective size is quantified by the convex-hull volume:
\begin{equation}
V(\theta)=\mathrm{Vol}\!\big(\mathrm{ConvHull}(\mathcal{W}_\theta)\big).
\end{equation}
This provides a conservative yet informative measure of workspace enlargement achievable by reconfiguring $\theta$.

\subsubsection{Gradient  for a Target Field}
In three–coil actuation for miniature helical robot control, we typically command only the rotating field’s magnitude and direction, while spatial gradients are not independently controllable and are therefore treated as disturbances (three inputs for six field/gradient components).  In this study, we characterize the disturbance gradients that arise under field control and examine how they can be exploited.

Given a target field of magnitude \(B_0\) along \(\mathbf d(\alpha,\varepsilon)\) at point $\mathbf p$, currents are synthesized via the least–norm solution. Using the precomputed per–coil gradient basis \(\{\mathcal G_k(\mathbf p,\theta)\}\), we evaluate the resulting gradient at \((\mathbf p,\theta)\) and define
\[
g_F(\alpha,\varepsilon,\mathbf p)\;=\;\big\|\nabla \mathbf B(\alpha,\varepsilon,\mathbf p)\big\|_{F}.
\]
By the threefold rotational symmetry of our coil layout about the system axis, \(g_F\) is expected to be invariant to azimuth \(\alpha\) for a fixed elevation \(\varepsilon\) due to the coil group's central symmetry. For a compact 1-D summary, we report the azimuth-averaged profile at each elevation
\[
M(\varepsilon,\mathbf p)\;=\;\frac{1}{2\pi}\int_{0}^{2\pi} g_F(\alpha,\varepsilon,\mathbf p)\,d\alpha,
\]
using uniform sampling of \(\alpha\) and a common \(B_0\) and current limits across all \(\theta\).

\subsubsection{Mean Force under a Rotating Field}
Small-scale helical robots driven by rotating fields typically operate at low frequencies in highly viscous media (about 3–10\,Hz). It is therefore appropriate to evaluate the cycle-averaged magnetic loading.

Assuming phase locking (\(\mathbf m \parallel \mathbf B\)) and rotation about axis \(\hat{\mathbf n}\), we command
\(\mathbf B(\phi)=B_0\,\hat{\mathbf b}(\phi)\) in the plane orthogonal to \(\hat{\mathbf n}\) and synthesize currents at each phase \(\phi\) using the same least-norm procedure as above. The dipole–gradient force satisfies
\[
\mathbf F(\phi)\ \propto\ \big(\nabla \mathbf B(\phi)\big)^{\!\top}\mathbf B(\phi),
\]
and we define the cycle average
\[
\bar{\mathbf F}(\hat{\mathbf n})=\frac{1}{2\pi}\int_{0}^{2\pi}\mathbf F(\phi)\,d\phi\, .
\]
For interpretation we report the vertical (lift) component
\[
F_z=\bar{\mathbf F}\cdot\hat{\mathbf z},
\]
\red{Besides \(F_z\), forces parallel and perpendicular to the rotation axis exist; however, in our mechanism they are not prominent relative to \(F_z\). A detailed analysis is provided in the project website. }


\begin{table*}[t]
\centering
 \captionsetup{font=footnotesize,labelsep=period}
\caption{Coordinate setpoints and depth references used in physical experiments. $\mathcal{F}_w$ denotes the world frame and $\mathcal{F}_m$ denotes the mechanism frame. Unless stated otherwise, the robot arm tracks the robot to keep $(x,y)\approx(0,0)$ in $\mathcal{F}_m$.}
\label{tab:exp_setpoints}
\begin{tabular}{p{2.2cm} p{4.7cm} p{5.0cm} p{4.8cm}}
\hline
Experiment & Trajectory / endpoints in $\mathcal{F}_w$ & Depth reference / windows (signal and thresholds) & Notes \\ \hline

Exp.~1: Straight line in tube &
Start $(0.295,0,0.03)$~m $\rightarrow$ End $(0.350,0,0.03)$~m &
Target depth in $\mathcal{F}_m$: $z=-0.23$~m (all $\theta$) &
Tube test at fixed depth; field magnitude ramped (see text / Fig.~\ref{fig:real_platform_exp}b). \\ \hline

Exp.~2: Circular cruise in free space &
Circle center $(0.295,0,0.034)$~m; radius $0.02$~m &
Commanded depth setpoints in $\mathcal{F}_m$:
$z=-0.24$~m (35$^\circ$),
$z=-0.22$~m (45$^\circ$),
$z=-0.20$~m (55$^\circ$) &
Mechanism lowered slightly to place the robot in the shallowest feasible layer for each $\theta$ while avoiding collision. \\ \hline

Exp.~3: 3D straight ramping in free space &
Start $(0.27,0,0.038)$~m $\rightarrow$ End $(0.31,0,0.048)$~m &
Depth windows in $\mathcal{F}_m$:
$z\in[-0.24,-0.23]$~m (35$^\circ$),
$z\in[-0.23,-0.22]$~m (45$^\circ$),
$z\in[-0.20,-0.19]$~m (55$^\circ$) &
Ramp executed within the shallowest feasible layer for each $\theta$ to compare maximum gradient/lift capability fairly. \\ \hline

Exp.~4: Depth-based $\theta$ scheduling in a tortuous tube &
Compound trajectory in a tortuous tube (see Fig. \ref{fig:theta_schedule_exp}a)&
Depth signal in $\mathcal{F}_w$: $z$ from miniature robot's position;  
common collision-free lower boundary: $z=0.075$~m in $\mathcal{F}_w$.&
Same trajectory executed with fixed $\theta\in\{35^\circ,45^\circ,55^\circ\}$ and scheduled $\theta(z)$; compare $E/B^2$ and $P(I_{\text{peak}}\ge 7~\mathrm{A})$. \\ \hline

\end{tabular}
\end{table*}

\section{Experiments and Results}
We evaluate DexterousMag using (i) simulation-based workspace analysis and (ii) four physical experiments that test the key trade-offs predicted by the simulations: depth reach, planar controllability, and gradient-assisted lift.
\subsection{Simulation Studies}
\red{We use an FEM-based pipeline to quantify how the polar angle~$\theta$ redistributes magnetic field energy, feasible workspace under current limits, and gradient-induced effects relevant to helical robot actuation.}

\subsubsection{Field-energy redistribution with depth}
\red{Fig.~\ref{fig:simulation res}a visualizes the magnetic energy density distribution (log-scale) under identical coil currents. As $\theta$ decreases, the energy distribution extends deeper along negative~$z$, indicating improved penetration. Along the symmetry axis (C\textsubscript{3}), Fig.~\ref{fig:simulation res}b further shows a depth-dependent crossover: larger~$\theta$ concentrates higher energy near the coil assembly, whereas smaller~$\theta$ sustains comparatively higher energy at greater depths. This confirms a depth-strength trade-off governed by $\theta$.}

\subsubsection{Feasible workspace under current limits}
\red{Under a per-coil current limit of 5~A, Fig.~\ref{fig:simulation res}c visualizes feasible workspaces $\mathcal{W}_\theta$ (colored by worst-case directional capability $B_{\min}$). Smaller $\theta$ produces a noticeably larger and deeper feasible region for generating a 1~mT rotating field. The workspace volume (Fig.~\ref{fig:simulation res}d) increases as $\theta$ decreases, demonstrating that $\theta$ can be used to trade peak capability for depth reach.}

\subsubsection{Gradient disturbance and mean lift}
\red{Fig.~\ref{fig:simulation res}e maps the induced gradient magnitude $g_F=\|\nabla\mathbf{B}\|_F$ while tracking a 1~mT target field. Across all $\theta$, $g_F$ decays with depth and increases when the field is closer to horizontal (small elevation angle). Importantly, larger~$\theta$ yields sharper near-coil gradient peaks, which can amplify undesired disturbance forces during planar control. Fig.~\ref{fig:simulation res}f reports the cycle-averaged vertical component, indicating that larger~$\theta$ produces stronger near-coil lift, whereas smaller~$\theta$ maintains non-negligible lift deeper in the workspace. These results motivate the four physical experiments below.}

\subsection{Physical Experiments}
\red{To validate the simulated trends, we conducted four physical experiments on the DexterousMag platform (Figs.~\ref{fig:real_platform_exp} and \ref{fig:theta_schedule_exp}). 
Experiments~1-3 benchmark depth reach, planar tracking, and gradient-assisted lift under three fixed mechanism angles $\theta\in\{35^\circ,45^\circ,55^\circ\}$. 
Experiment~4 extends the above tests to a constrained, more realistic scenario: the robot traverses a tortuous tube that concatenates the three task primitives, while an acrylic plate is placed above the workspace to emulate anatomical thickness and impose a collision constraint. Using the robotic arm for global re-positioning, we align a \emph{collision-free lower boundary} across configurations and then compare fixed $\theta\in\{35^\circ,45^\circ,55^\circ\}$ against the depth-based $\theta$ scheduled mode along the same compound trajectory.}

\subsubsection{Experimental setup}
The miniature helical robot (Fig.~\ref{fig:concept}) follows the design of Xu \textit{et al.}~\cite{xu_image-based_2020} and was tested in 350~cSt silicone oil to emulate low-Reynolds-number conditions. A robot arm maintained the robot near the C\textsubscript{3} axis to reduce asymmetry. Each coil was current-limited to 5~A and we performed five trials per condition ($n{=}5$). \red{The detailed coordinate setpoints and depth references used in experiments is summarized in Table \ref{tab:exp_setpoints}.}

\paragraph{Experiment 1: Depth reach in a tube.}
\red{We evaluated depth reach by driving the robot along a straight segment in a tube while ramping the commanded field magnitude. The key metric is the total current required to reach the onset of sustained forward motion at the target depth. As shown in Fig.~\ref{fig:real_platform_exp}b, smaller $\theta$ required substantially less current to achieve the same transition, whereas $\theta=55^\circ$ reached the current limit earlier and exhibited loss of synchrony at depth, consistent with the simulated feasible-workspace trends.}

\paragraph{Experiment 2: Planar tracking accuracy versus gradient disturbance.}
\red{We assessed planar controllability by tracking a circular path at fixed field magnitude ($B_0=1$~mT) using the same controller across $\theta$. Figure~\ref{fig:real_platform_exp}c shows that $\theta\approx45^\circ$ achieved the lowest tracking error with the lowest current cost, while $\theta=55^\circ$ degraded both accuracy and efficiency, consistent with larger near-coil gradients inducing stronger  force disturbances.}

\paragraph{Experiment 3: Gradient-assisted lift in $z$-ramping.}
\red{We tested gradient-assisted lift by commanding a straight $z$-ramping trajectory. As shown in Fig.~\ref{fig:real_platform_exp}d, $\theta=55^\circ$ produced the most monotonic and steep $z$ progression, while smaller $\theta$ yielded weaker lift over the same duration. The improved lift at $\theta=55^\circ$ did not require higher current cost, indicating that concentrating gradient capability near the coils benefits tasks that rely on vertical force components.}

\paragraph{Experiment 4: Depth-based $\theta$ scheduling on a compound trajectory.}
\red{Motivated by Experiments~1-3, we evaluated a depth-triggered deterministic \(\theta\) schedule along a compound trajectory spanning three depth regimes.(Fig.~\ref{fig:theta_schedule_exp}a, b):}
\[
\theta(z)=
\begin{cases}
35^\circ, & z<0.035~\mathrm{m}\quad \text{(deep/depth reach)},\\
45^\circ, & 0.035<z<0.045~\mathrm{m}\quad \text{(mid/cruise)},\\
55^\circ, & z>0.045~\mathrm{m}\quad \text{(shallow/ramping)}.
\end{cases}
\]
\red{For fair comparison, we aligned the lower feasible boundary across configurations (to $z=0.075$~m in $\mathcal{F}_w$) and repeated the same trajectory with fixed $\theta\in\{35^\circ,45^\circ,55^\circ\}$ and with the scheduled mode.}

\red{Figure~\ref{fig:theta_schedule_exp}c reports the energy proxy $E/B^2$ and the high-current risk $P(I_{\text{peak}}\ge 7~\mathrm{A})$ over the full run and within each depth regime. Energy proxy $E/B^2$ is defined as the ratio of mean squared current to the square of the mean magnetic field magnitude per 1-second window. This normalized index evaluates the current cost relative to the generated field strength.  $I_{\text{peak}}=\max_{t\in \mathrm{window}}\|\mathbf I(t)\|_2$ is computed within each 1~s window; note that $I_{\text{peak}}$ may exceed 5~A because it is the norm of the three-coil current vector, while each coil is individually saturated at 5~A. 
We prioritize $P(I_{\text{peak}}\ge 7~\mathrm{A})$ as a key performance indicator to address a critical trade-off: although increasing current allows for field generation at any depth, it triggers low energy efficiency, parasitic vibrations via inter-coil interaction forces and gradient fluctuations. DexterousMag distinguishes itself by prioritizing an operational regime that circumvents these high-current hazards, fostering an electromagnetic system that is both energy-efficient and structurally stable. 
Over the full trajectory, scheduled $\theta$ reduced high-current events to \textbf{2.1\%}, while keeping $E/B^2$ competitive.
Across regimes, the schedule avoids the dominant failure modes of fixed angles: in the shallow/climb regime, fixed 35$^\circ$ exceeded the peak-current threshold in all windows, whereas the scheduled mode and fixed 55$^\circ$ avoided such events; in the mid/cruise regime, fixed 45$^\circ$ achieved the lowest $E/B^2$, consistent with energy-efficient cruising (and under constrained motion, reduced $z$-regulation can make 55$^\circ$ appear similarly efficient); and in the deep regime, fixed 55$^\circ$ exhibited frequent margin violations, whereas the scheduled mode and fixed 35$^\circ$/45$^\circ$ remained within the threshold (\textbf{0\%}).}

\red{Overall, a simple depth-based schedule leverages reconfigurability to improve task-level robustness under current constraints, combining the deep-workspace margin of smaller $\theta$, the cruising efficiency near $\theta\approx45^\circ$, and the shallow/climb safety of larger $\theta$.}

\section{Conclusion}
This letter presents \textit{DexterousMag}, a robot-arm-assisted three-coil electromagnetic actuation system whose compact parallel mechanism exposes the coil-group polar angle \(\theta\) as a reconfigurable geometric DOF. We precompute FEM-based libraries \(\mathbf{A}(\mathbf{p},\theta)\) and \(\mathcal{G}_k(\mathbf{p},\theta)\) to quantify field/gradient capability and feasibility under current and collision constraints, revealing a depth-strength trade-off across \(\theta\). Experiments validate these predictions, and an added depth-triggered, piecewise \(\theta(z)\) schedule demonstrates online reconfiguration on a compound tube task for improved robustness. 
\red{A limitation of the present prototype is its restricted coil size for clinical translation. Nevertheless, the role of \(\theta\) is expected to persist under scaling: \(\theta\) changes the coil-group geometry and thus redistributes field/gradient authority in a manner consistent with standard scaling behaviour, so the same trade-offs and scheduling principle should apply when integrated with larger coils and higher-payload clinical manipulators.}
\bibliographystyle{IEEEtran}
\bibliography{references_zotero}

\vspace{12pt}

\end{document}